# Optimizing Transistor Performance of Percolating Carbon Nanotube Networks

V. K. Sangwan[1,2*], A. Behnam[3], V. W. Ballarotto[2], M. S. Fuhrer[1], A. Ural[3] and E. D. Williams[1,2]

[1] Department of Physics, University of Maryland, College Park, Maryland 20742

[2] Laboratory for Physical Sciences, College Park, Maryland 20740

[3] Department of Electrical and Computer Engineering, University of Florida, Gainesville, Florida 32611

**Abstract:**

In percolating networks of mixed metallic and semiconducting CNTs, there is a tradeoff between high on-current (dense networks) and high on/off ratio (sparse networks in which the metallic CNT fraction is not percolating). Experiments on devices in a transistor configuration and Monte Carlo simulations were performed to determine the scaling behavior of device resistivity as a function of channel length ($L$) for CNT density $p$ in the range 0.04 - 1.29 CNT/$\mu m^2$ in the on- and off-states. Optimized devices with field-effect mobility up to 50 cm$^2$/Vs at on/off ratio > $10^3$ were obtained at $W = 50$ μm, $L > 70$ μm for $p = 0.54 – 0.81$ CNTs/$\mu m^2$.

* Current address: Materials Science and Engineering, Northwestern University, IL 60208

Recently, random networks of carbon nanotubes (CNTs) as well as aligned arrays of CNTs have been demonstrated as potential active materials in large-area electronics applications [1-3]. Although aligned arrays of CNTs can carry large channel current, they are plagued by low on/off ratio due to metallic CNTs directly shorting the electrodes [4]. Metallic paths can be eliminated by electrical breakdown method [3, 5], but this has proved difficult to scale up for large area integrated electronics. In contrast, the effects of metallic CNTs in a random network can be mitigated by carefully controlling the CNT density and device geometry such that the metallic fraction of the CNTs is below the percolation threshold (i.e. every conducting path contains at least one semiconducting CNT) [1, 6]. This strategy contains an inherent tradeoff: high device current and field-effect mobility can be achieved by increasing the number of gate-tunable paths in the channel, i.e. by increasing CNT density; however, a very high CNT density results in reduced on/off ratio due to increased number of metallic CNTs. An optimized device (i.e. highest possible on-current at a given on-off ratio) has the total density of CNTs above the percolation threshold, but the density of metallic CNTs below the percolation threshold. Although such high quality devices have been reported in the literature [1, 2], a controlled and systematic experiment to determine the CNT density for optimum device performance is lacking. Here, we study the effect of device parameters (density of CNTs $p$ and channel length $L$) on transport properties to obtain optimized device performance. The CNT density $p$ was varied over 2 orders of magnitude by carefully controlling the chemical vapor deposition process. For each density, scaling behavior of device resistance as a function of channel length ($R \sim L^n$) was demonstrated and used to determine the overall percolation threshold and the percolation threshold of the metallic CNTs. We also performed Monte Carlo simulation of random networks of CNTs to confirm the observed percolative behavior of CNT thin films. Field-effect mobility (~ 50 cm$^2$/Vs) and on/off ratio (~ 4 x 10$^3$) of the optimized devices are comparable to the best devices reported in the literature [1].

CNT thin films were grown by CVD on 300 nm thick thermally oxidized Si substrates using Fe as the catalyst [7, 8]. As described elsewhere [8], the density of the CNT network was controlled in the range 0.04 – 1.29 CNTs/µm$^2$ by varying the concentration of ferric nitrate catalyst. The density of CNTs was determined by counting the number of CNTs in field-emission scanning electron microscope (FE-SEM) images. CNTs were counted in 40 µm x 60 µm rectangles (magnification = 5 kX) for sparse CNTs thin films ($p \leq 0.57$ CNTs/ µm$^2$) and in 12 µm x 16 µm rectangles (magnification = 20 kX) for denser CNT thin films ($p > 0.57$ CNTs/ µm$^2$). The root-mean square (RMS) length of the CNTs ($l_{CNT}$) was determined to be approximately 5 µm. Source-drain electrodes (100 nm thick Au on 10 nm thick Ti) were deposited on growth substrate by photolithography. Then, the CNT thin films were patterned in the device channel via another step of photolithography and reactive ion etching [8] to obtain channel width $W = 50$ µm and channel lengths $L$ varying from 5 µm to 100 µm in steps of 5 µm (20 devices in total). The resulting devices consist of a highly doped Si as global back gate, 300 nm thick SiO$_2$ dielectric layer and top contact Ti/Au S/D electrodes.

Transport measurements of the devices were conducted in ambient conditions using a probe station (Cascade Microtech). Devices were characterized by computing device resistance, linear effective field-effect mobility (the field effect mobility assuming that the device is a uniform thin film of material) and on/off ratio. We calculated device resistance in the ON-state ($R_{on}$ at $V_g = -30$ V) and OFF-state ($R_{off}$ at $V_g = 30$ V) from the inverse slope of the $I_d$-$V_d$ characteristics in the linear regime $-1\,\text{V} < V_d < 1\,\text{V}$; $V_g$ is gate voltage, $V_d$ is drain voltage, and $I_d$ is drain current. The effective field-effect mobility $\mu = \dfrac{L}{V_d C_g}\left|\dfrac{\partial I_d}{\partial V_g}\right|$ is measured in the linear regime of $I_d(V_d)$. We studied the transport behavior as a function of $L$ for 10 different CNT densities in the range $p = 0.04 - 1.29$ CNTs/μm$^2$. A total of 170 devices were measured (20 devices for each of CNT densities: $p = 0.04, 0.16, 0.23, 0.34, 0.57, 0.63$ and $0.81$ CNTs/μm$^2$, and 10 devices for each of CNT densities: $p = 1.09, 1.15$ and $1.29$ CNTs/μm$^2$).

The transport properties of CNT TFTs have interpreted within the framework of percolation theory of a random network of conducting sticks [9, 10]. Though scaling laws in percolation theory are valid for infinite sized systems, they offer approximate solution of the finite devices considered here. For $W \gg L$ and $L \gg l_{CNT}$, the resistance ($R_s$) of a conducting stick thin film is expected to follow a power law relation with channel length $L$ [2, 11]. At the percolation threshold, $R_s$ varies as $\sim L^{1.8}$, and at a density much above the percolation threshold, $R_s$ varies linearly with $L$ (approaching a homogeneous film) [2, 9-11]. In between these two extremes, the exact form of the power law depends on the conducting stick density. Factors which might affect the form of power law observed experimentally include heterogeneity in the electronic type of CNTs (1/3 metallic and 2/3 semiconducting), CNT-to-CNT contact resistance [12], twisted CNTs instead of straight stick assumed in the model, and variations in CNT length.

Device resistance ($R$) is plotted against $L$ as a log-log plot to extract the scaling exponent as illustrated in Fig. 1. In the ON-state, all the CNTs conduct, therefore the ON-state exponent, $n_{on}$ ($R_{on} \sim L^{n_{on}}$) corresponds to the total density of CNTs. In the OFF-state, only metallic CNTs carry current, therefore $n_{off}$ ($R_{off} \sim L^{n_{off}}$) is expected to give information about the density of metallic CNTs. For the smallest CNT density $p = 0.04$ CNTs/μm², only the smallest channel length device ($L = 5$ μm) showed a conducting channel with $R \sim 100$ kΩ and on/off ratio ~ 2. In this case, CNT thin film is too sparse to form a connected network and the conducting channel is formed by CNTs directly shorting source-drain electrodes. At higher CNT density, bigger clusters of connected CNTs result in more and more devices with conducting channels. For example, at $p = 0.16$ and $0.23$ CNTs/μm², conducting devices were obtained up to $L = 10$ and $25$ μm, respectively. At $p = 0.34$ CNTs/μm², all the devices up to $L = 100$ μm showed conducting channels with $n_{on} = 2.19$, which suggests that density of CNTs is approaching the percolation threshold. For $p = 0.57$ CNTs/μm², $R$ and field-effect mobility ($\mu$) is plotted against $L$ in Fig. 1(a) and (b), respectively. At this density, CNT TFTs showed field-effect mobility between $1 - 50$ cm²/Vs and on/off ratio in range $2 - 10^4$. In this case, ON-state exponent $n_{on} = 1.59$ and the OFF-state exponent $n_{off} = 4.21$. Both the exponents, $n_{on}$ and $n_{off}$, decrease with an increase in CNT density (Fig. 1(a), (c) and (e)). The on/off ratio ($I_{on}/I_{off} = R_{off}/R_{on}$) of the devices increases with $L$ (Fig. 1(a)) whereas, field-effect mobility decreases with $L$ (Fig. 1(b)). However, dependence of on/off ratio and field-effect mobility on $L$ becomes weaker as CNT density is increased (Fig. 1(e) and (f)). The observation of exponents $n_{off}$ larger than 1.8 suggest only that metallic CNT density is below the percolation threshold, without showing the extent of percolation in the network. In addition, the scaling behavior in the OFF-state resistance could be affected by the non-perfect turn-off of some semiconducting CNTs and the presence of a few long CNTs in the channel.

The average field-effect mobility and the ON-state current both increase with CNT density. In contrast, the mean on/off ratio of the devices decreases with increasing network density (Fig. 1(a), (c) and (e)), even though the on/off ratio of the devices with longer channel lengths ($L > 70$ μm) remains higher than $10^3$ for $p \leq 0.81$ CNTs/μm$^2$. This means metallic CNT clusters are not large enough to bridge long channels ($L > 70$ μm) for $p \leq 0.81$ CNTs/μm$^2$. At $p = 1.09$ CNTs/μm$^2$, the on/off ratio suddenly drops to the range $4 - 40$ and the OFF-state exponent becomes $n_{off} = 1.95$ (Fig. 1(e)). This suggests that density of metallic CNTs in the network has become close to the percolation threshold. For $p > 1.09$ CNTs/μm$^2$, on/off ratio decreases further and both the exponents $n_{on}$ and $n_{off}$ approach 1. At the highest CNT density $p = 1.29$ CNTs/μm$^2$, the field-effect mobility of the devices is observed to be between $50 - 190$ cm$^2$/Vs, and on/off ratio between 2 to 6.

Monte Carlo simulations were performed to quantify the scaling behavior of device resistance, following the method described previously in Ref. [13]. Random networks of 5 μm long CNTs (RMS length of CNTs in experiments [14]) were generated in an area of fixed width and variable channel length, as shown in Fig. 2(a). It was observed that scaling behavior in these device geometries becomes almost independent of $W$ for $W > 15$ μm ($W >> l_{CNT}$) [13]. Therefore, $W$ was fixed at 25 μm due to computational limitations. First, the percolation threshold was determined by calculating the connection probability (defined as the ratio of devices with connected source-drain electrodes to total number of simulated devices) as a function of CNT density. Ensemble of 200 devices were simulated for each CNT density in a $W = 25$ μm x $L = 35$ μm channel area. The percolation threshold, which we define as the density at which the connection probability is 0.99 was found to be 0.28 CNTs/μm$^2$, (Fig. 2(b)). This simulated percolation threshold is close to the minimum CNT density (0.34 CNTs/µm$^2$) where all the devices (up to $L$ = 100 µm) showed a connected channel. Next, resistances ($R$) of the simulated networks were calculated based on the resistance values assigned to CNTs and their junctions [13]. Fig. 2(c) shows $R$ as a function of $L$ for CNT densities varying from 0.2 to 0.8 CNTs/µm$^2$ (numbered 1 to 8). Solid lines are least squares fits to extract scaling exponents which decreased from above 2 to approximately 1.5 with increasing density.

In Fig. 3(a), the scaling behavior of device resistance in experiment and Monte Carlo simulations is summarized by plotting the experimental exponents ($n_{on}$ and $n_{off}$) as well as simulated exponent ($n_{on}$) against the CNT density. In the density range of $p$ = 0.57 – 0.81 CNTs/µm$^2$, $n_{on}$ < 1.8 and $n_{off}$ > 1.8; the overall CNT density is above the percolation threshold, whereas the density of metallic CNTs is below the percolation threshold. Devices with optimum performance (field-effect mobility ~ 10 – 50 cm$^2$/Vs, on/off ratio > 10$^3$) are obtained at $L$ > 70 µm and $p$ = 0.34 – 0.81 CNTs/µm$^2$. The highest field-effect mobility, for the devices with on/off ratio > 10$^3$, is obtained at $p$ = 0.81 CNTs/µm$^2$, which is the highest density CNTs where the density of metallic CNTs is still below the percolation threshold. The field-effect mobility (up to 50 cm$^2$/Vs) and on/off ratio (up to 4 x 10$^3$) of these devices is comparable to one of the highest quality devices reported in literature [1]), where the on/off ratio was improved by limiting metallic paths via lateral confinement of CNTs in long and narrow stripes. Here, comparable device performance was achieved without utilizing lateral confinement [1] or electrical breakdown [5] to minimize the effect of metallic CNTs. Presumably such additional processing could further enhance the on/off ratio. The effect of percolation on device performance can be seen more clearly by considering devices with $L$ = 100 µm. In Fig. 3(b), on/off ratio and ON-state current ($I_{on}$) of the device is plotted as a function of CNT density. At $p < 0.81$ CNTs/μm$^2$, on/off ratio of the device remains in between $10^3 - 10^4$, whereas device current of the device increases steadily with CNT density. Above $p = 1.0$ CNTs/μm$^2$, on/off ratio drops sharply while device current continues to increase. The optimum range of CNT density to achieve high mobility while maintaining and on/off ratio $> 10^3$ is approximately $0.5 - 1.0$ CNTs/μm$^2$.

In conclusion, we performed controlled and systematic experiments and simulations to obtain the range of device parameters (CNT density *p* and channel length *L*) for optimized performance of CNT TFTs. Using this approach, CNT TFTs with field-effect mobility between $5 - 50$ cm$^2$/Vs and on/off ratio$> 10^3$ were obtained at $W = 50$ μm, $L > 70$ μm and $p = 0.54 - 0.81$ CNTs/μm$^2$.

**Acknowledgements:**

This work has been supported by the Laboratory for Physical Sciences, and by use of the UMD-MRSEC Shared Equipment Facilities under Grant No. DMR 05-20471. Infrastructure support is also provided by the UMD NanoCenter and CNAM.8

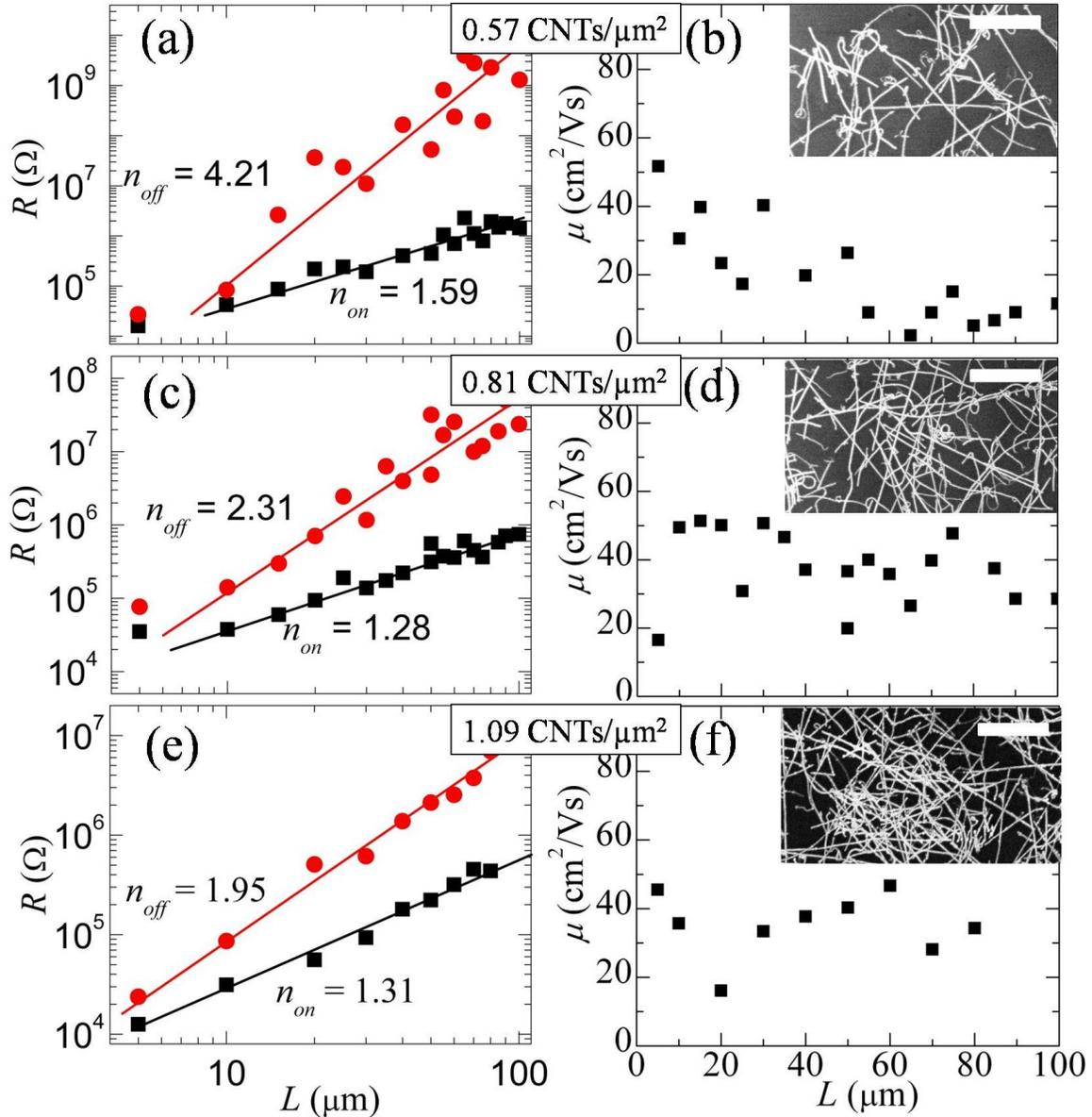

Figure 1. (a), (c) and (e) Device resistance (R) in ON-state (black squares) and OFF-state (red circles) versus channel length (*L*) as CNT network density (shown in the box in the center) is increased from top to bottom. (b), (d) and (f) Field-effect mobility (*μ*) versus *L* for different densities of CNTs. Insets in (b), (d) and (f) show FE-SEM images of corresponding CNT densities. White scale bars are 2 μm long.

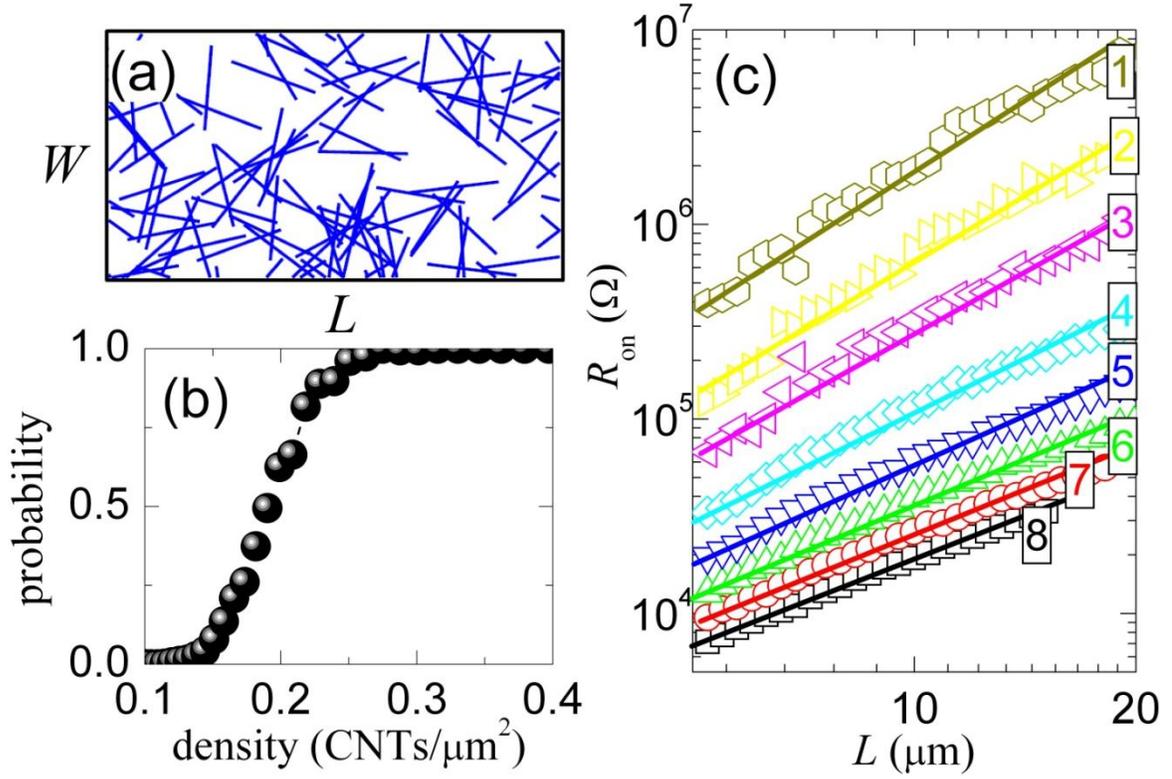

Figure 2. (a) A 2D CNT network generated using Monte Carlo simulations. (b) Probability of connection between source and drain is plotted as a function of CNT density for $W = 25$ μm, $L = 35$ μm and $l_{CNT} = 5$ μm. The percolation threshold is defined as probability = 0.99. (c) Simulated resistance of CNT devices is plotted against $L$ for eight different CNT densities varying from 0.2 (#1) to 0.8 CNTs/μm$^2$ (#8).

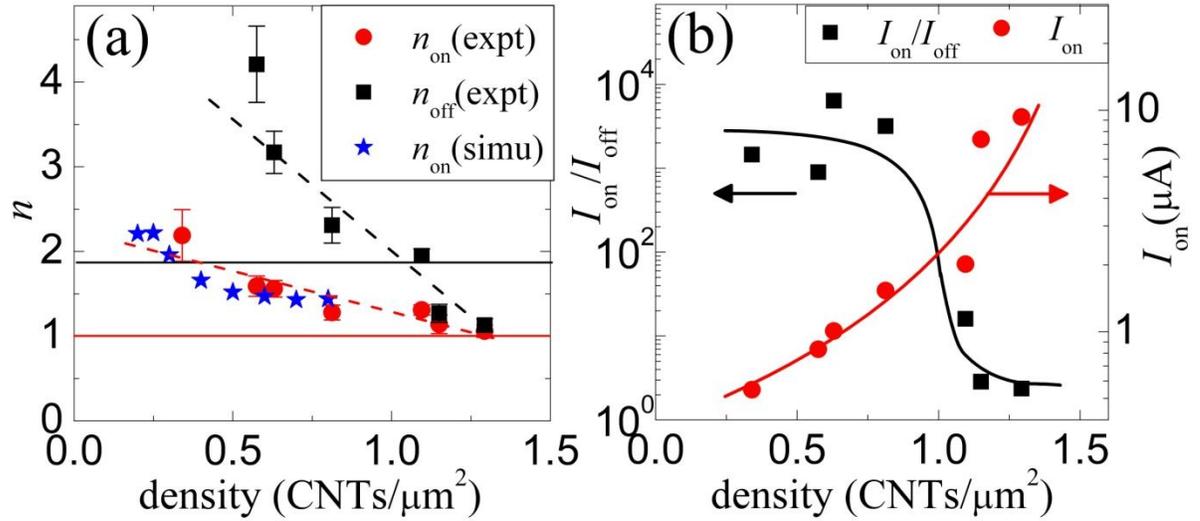

Figure 3. (a) Resistance power law exponents in ON- and OFF-states are plotted as a function of CNT density. Horizontal black line at $n = 1.8$ shows the percolation threshold, whereas horizontal red line at $n = 1$ represents the limit for a homogeneous film. The blue stars represent simulated ON-state exponents. Slanted dashed lines are to aid the eye. (b) On/off ratio and ON-state current are plotted as a function of CNT density for a device with $L = 100$ μm and $W = 50$ μm. Curved solid lines are to aid the eye.